\documentstyle[epsf,twocolumn,prl,aps]{revtex}

\newcommand{\tj}{$t$-$J$\ }

\newcommand{\etal}{{\it et al}}
\newcommand{\ij}{\langle ij\rangle}
\renewcommand{\S}{{\vec S}}  
\newcommand{\grapprox}{\stackrel{>}{_\sim}}

\begin{document}
\draft

\twocolumn[\hsize\textwidth\columnwidth\hsize\csname @twocolumnfalse\endcsname

\title{Phase separation and stripe formation in the 2D \tj
model: a comparison of numerical results}
\author{ Steven R.\ White$^1$ and D.J.\ Scalapino$^2$}
\address{ 
$^1$Department of Physics and Astronomy,
University of California,
Irvine, CA 92697
}
\address{ 
$^2$Department of Physics,
University of California,
Santa Barbara, CA 93106
}
\date{\today}
\maketitle
\begin{abstract}
\noindent 

We make a critical analysis of numerical results for and against
phase separation and stripe formation in the \tj model.
We argue that the frustrated phase separation mechanism for
stripe formation requires phase separation at too high a doping
for it to be consistent with existing numerical
studies of the \tj model. 
We compare variational energies for various methods, and conclude that 
the most accurate
calculations for large systems appear to be from the density
matrix renormalization group. These calculations imply that the
ground state of the doped \tj model is striped, not phase separated.

\end{abstract}

 ]

The recent discovery of stripes in the underdoped
cuprates \cite{stripeexper} has
brought renewed interest to the question of the existence
of phase separation in the \tj and Hubbard models. Interest
in this question began a decade ago, when evidence for
phase separation in La$_2$CuO$_{4+\delta}$ was found \cite{jorgensen}.
In particular, Emery, et. al.
argued that rather than stemming from the chemistry of the
mobile oxygen atoms in this material, 
the phase separation reflected a universal 
tendency for doped holes in antiferromagnets to phase
separate \cite{EKL}. They argued that this phase separation is
not seen in the absence of mobile dopants because it is
frustrated by the long-range Coulomb repulsion between holes.
This led to a number of studies using analytical arguments
and numerical simulations to find out whether the \tj and
Hubbard models (which do not have long-range Coulomb interactions) 
exhibit phase separation \cite{earlyphasesep} and to the related question of
the mechanism responsible for stripe formation in the cuprates.

There are, in fact, currently two main views regarding the origin of stripes.
In the first, stripes form because of a competition between
kinetic and exchange energies in doped antiferromagnets.
In this approach, long range Coulomb interations are not
important.
Indeed, a decade ago Hartree-Fock solutions of the 2D Hubbard 
model showed that domain walls were present in mean field
solutions of the Hubbard model \cite{hfdomain}.
However, the stripes in the Hartree-Fock solution are
characterized by a filling of one hole per domain wall unit
cell, while the incommensurate spin susceptibility
peaks seen in experiments require a filling of half this. 
Subsequently, numerical studies of the \tj model
by Prelovsek and coworkers \cite{prelovsek} 
showed that indeed, stripe-like correlations
are an important ingredient in the ground state
of small \tj clusters. This work also made clear that stripes
act as domain walls in N\'eel antiferromagnets. However,
because of the limited size of the systems studied, only
filled stripes were found.
Recently, using density matrix renormalization group
methods (DMRG) \cite{dmrg} to study much larger systems numerically, 
we have found evidence for
striped ground states for a wide range of dopings in the \tj
model \cite{stripe,energetics}.  Significantly, we have found that
stripes with a linear doping of one hole per two domain wall
unit cells are the lowest energy 
configurations at low doping.

The second approach starts with the assumption that without
long-range Coulomb interactions, doped antiferromagnets
phase separate. Stripe formation arises in this approach
because the long-range 
Coulomb repulsion frustrates the phase separation,
leading to an inhomogeneous charge density state \cite{topologicaldoping}. 
The $\pi$ phase
shift characteristic of a domain wall arises in this
``frustrated phase separation'' approach
from a secondary effect, namely from the same reduction of the
transverse kinetic energy of
hopping which {\it drives} domain wall formation in the
first approach.  In support of this approach, studies of
classical spin models of competing long and short range interactions
have been shown to have a variety of inhomogeneous ground states, including
striped states \cite{low}.  Unfortunately, the difficulty associated with
the long-range Coulomb interactions has so far prevented more realistic
microscopic calculations. However, as a minimum requirement for
the viability of the frustrated phase separation scenario,
one clearly must have phase separation in relevant models
of doped antiferromagnets which lack the long-range Coulomb interaction, 
such as the \tj or Hubbard models. Extensive numerical studies of the
Hubbard model have failed to find convincing evidence of 
phase separation \cite{earlyphasesep} and
interest has shifted to the \tj model which does exhibit phase separation
in certain regions of $J/t$-doping parameter space. In this case, the
question becomes one of determining whether the phase-separation takes
place in the physical parameter range. More generally, the question becomes
one of whether more elaborate models such as, for example, 3-band models or
models which include electron-phonon interactions will exhibit phase
separation in the physical region of parameter space.  Here we will not
address this more general question, but rather focus on the \tj model
because it has often been used in the discussion of stripe formation.

The proposal that the doped \tj model phase-separates in
physically relevant parameter and doping regimes has been supported by
variational arguments \cite{EKL}, diagonalizations of small
clusters \cite{EKL},
Green's function quantum Monte Carlo (QMC) calculations \cite{hellberg}, 
and the recent
analysis of the Casimir force arising from fluctuating spin waves in
the antiferromagnetic regions
separating widely spaced stripes \cite{pryadko}. 
On the other hand, a substantial
body of other QMC calculations \cite{nophasesep}, series
expansions \cite{putikka},
exact diagonalizations \cite{dagotto}, 
and our DMRG calculations  \cite{stripe,energetics}
have yielded results contradicting these claims.
In this paper, we will review some of these
calculations and arguments. However, we will pose a slightly
different, and easier question regarding phase separation
than has generally been addressed in previous studies. 
Rather than asking, ``Does the \tj model phase separate at 
arbitrarily low doping?'', we will ask, ``Does the \tj model
phase separate at high enough doping to allow the frustrated
phase separation mechanism to yield stripes consistent with
those found in the cuprates?''. We will conclude that the answer
to this question is that it does not.

We will also compare
the variational energies of several of the numerical 
approaches. In this comparison, we find that the DMRG
calculations yield energies in excellent agreement with
exact diagonalization, but can be extended to much larger systems. 
On the other hand, we find that the best Green's function QMC
calculations to date are slightly higher in energy, and that
this energy difference is close to the stabilization energy of stripes over
pairs.  Our DMRG
calculations give striped ground states directly, 
without long range Coulomb interactions included in the model
and without phase separation.
Note that formation of a uniform array of stripes
is {\it not}  phase separation. 

Most of the numerical work on doped antiferromagnets has centered on the 
\tj model, with a Hamiltonian given by
\begin{equation}
H = - t \sum_{\langle ij \rangle s}
      ( c_{is}^{\dagger}c_{js}
                + {\rm h.c.}) + 
J \sum_{\langle ij \rangle}
      ( {\bf S}_{i} \! \cdot \! {\bf S}_{j} -
         \frac{n_i n_j}{4} ) ,
\label{tj-ham}
\end{equation}
where doubly occupied sites are explicitly excluded from the Hilbert
space.
Here $\ij$ are near-neighbor sites, $s$ is a spin index, $\S_i$
and $c^\dagger_{i,s}$ are electron spin and creation operators,
and $n_i= c^\dagger_{i\uparrow}c_{i\uparrow} +
c^\dagger_{i\downarrow}c_{i\downarrow}$.
The near-neighbor hopping and exchange interactions are $t$ and $J$.  
We measure energies in units of $t$. 

We begin with a review of previous arguments
and numerical data concerning phase separation.
Emery, Kivelson, and Lin \cite{EKL} used a combination
of variational arguments for large and small $J/t$
and exact diagonalization for moderate values of $J/t$ to argue
for the occurence of phase separation at all values of $J/t$.
The variational arguments show that for small enough $J/t$ 
(roughly $J/t < 10^{-2}$), and low enough doping,
a uniform paramagnetic phase of independent holes has higher energy than
a phase-separated state in which the hole-rich state is 
ferromagnetic.  Emery, et. al. pointed out that
this variational argument would not rule out other phase-separated 
states which might have even lower energy and be more physical
than the ferromagnetic state. However, these hypothetical
states might also be uniform \cite{putikka}. In particular,
Putikka, et. al. suggested \cite{putikka,ioffe} that a
uniform ferrimagnetic state might have lower energy than
the phase-separated ferromagnetic state. 
In any case, the
``Nagaoka-like'' ferromagnetic state which was shown to
have low energy is of limited relevance to the cuprates.

The exact diagonalization of Emery, et. al. showed, at $J/t=0.1$ and $0.4$, 
that two holes on a $4\times4$
system bind into a pair, but that two pairs do not bind.
Emery, et. al. argued that the pair formation was a sign
of phase separation at low doping. However, others argued that
the binding of two pairs, rather than pair formation, signals
the onset of phase separation \cite{dagottoonset}. The possibility
of striped ground states makes even the binding of two pairs
an unreliable indicator of phase separation: the two pairs
may form a short ``stripe'', but if stripes do not attract,
there is no phase separation.

Larger systems have subsequently been studied using several different
types of QMC and related techniques.
Almost all of these studies concluded that there was no
phase separation in the parameter regimes relevant to the 
cuprates \cite{nophasesep,putikka,dagottoonset,shih,calandra}. 
In contrast, using Green's function Monte Carlo,
Hellberg and Manousakis (HM) \cite{hellberg} concluded that
phase separation occurs at all values of $J/t$ at low enough
doping, and in particular that for $J/t=0.3$ it 
occurs for hole-doping levels less than about $x \sim 0.12$.  
Although in these various studies the possibility of striped ground states 
was usually not considered, and thus was not specifically excluded, 
the results obtained were generally uniform.
Our DMRG calculations represent a third possibility,
namely that stripes form in the $t$-$J$ model, 
without any phase separation into uniform hole-rich and 
undoped regions, and without the need to add long-range Coulomb terms. 
Consequently, there are three broad possibilities for the 
charge ordering of the ground state of the lightly-doped, 
pure $t$-$J$ model at $J/t \approx 0.3$: 
phase separated, uniform, or striped.

The question of what happens in the limit of very low doping
is quite difficult for numerical methods, requiring
increasingly larger systems as the doping is reduced.
Most of the controversy has centered on the very low doping
range.
Fortunately, if one is interested in the mechanism of
stripe formation at dopings $x$ 
relevant to the cuprates, say, from
$0.07 \le x \le 0.25$, one need not be concerned with extremely
low doping. The frustrated phase separation scenario, in
fact, appears to put rather strong constraints on the dopings
required to produce phase separation. First, note that according
to this scenario, phase
separation must occur at all dopings in which stripes are
found. Furthermore, note that the long-range part of the Coulomb 
interaction between holes makes the hole density distribution
more uniform.
Consider, therefore, as in Ref. \cite{EKL}, a phase-separated system, 
in the absence of long-range Coulomb forces, 
which has all of the holes in one region at a density 
$x_{\rm ps}$ and no holes in the other region. Then, turning on
the Coulomb interaction will tend to drive the holes apart,
possibly elongating the hole-rich regions into stripes,
if the long-range repulsion is not too weak and not too strong. 
Under these circumstances, the local hole density in the
stripes, $x_s$,
is lower, or, at most, the same, as the original hole density $x_{\rm
ps}$: $x_s \le x_{\rm ps}$.
If this is true, then the known lower limits for $x_s$ from
experiments imply lower limits on $x_{\rm ps}$.
Note that $x_{\rm ps}$ is simply the critical doping at which
phase separation first occurs, for any particular value of $J/t$.

Neutron scattering shows that the hole doping per unit length 
of the stripes for $x \le 0.125$ is about 0.5. The neutron
scattering experiments currently
cannot determine whether the stripes are centered on copper
sites (site-centered) or
on the oxygen sites between them (bond-centered). 
If one assumes the stripes
are site-centered, with nominal width 1 in the $t$-$J$ model,
then the local doping {\it within}  a stripe is $x_s=0.5$. If one
takes them as bond-centered, with nominal width 2, then
the doping within a stripe is $x_s=0.25$. Of course, the hole
density is not strictly zero in the antiferromagnetic regions between
stripes. On the other hand, there are some signs of stripes well
above $x=0.125$, and one would expect some suppression of
$x_s$ relative to $x_{\rm ps}$. Furthermore, if stripes are
necessary for superconductivity, one certainly needs them for
the whole superconducting doping range. Therefore, for the sake
of argument, we will assume that experiments require
$x_{\rm ps} \grapprox 0.25$. However, our conclusions would
not change if a somewhat smaller limit (say 
$x_{\rm ps} \grapprox 0.2$) were used.

Although there is disagreement about the behavior at smaller
doping, there is general agreement among the various approaches
on the lack of phase separation at $x=0.25$. For example,
HM report the critical doping level for $J/t = 0.3$ to be
about 0.12(2), implying the density of the hole-rich regions
is also $0.12(2)$, far from $x=0.25$. Kivelson, et. al.
interpretted the lack of binding of two pairs in \cite{EKL}
as indicating the critical doping was less than
$x=3/16=0.1875$ \cite{kivelson}.
We are not aware of any interpretations of 
quantitative numerical calculations
finding phase separation above this value.
What value of $J/t$ can give phase separation near $x=0.25$?
According to HM, one would need 
$J/t \sim 0.9$.
Other calculations find higher values of $J/t$. 
For example, Calandra, Becca, and Sorella
find $J/t \sim 1.0$, and series expansion techniques
produce an even higher value---$J/t \sim 1.5$.

Although the calculations seem in reasonable agreement 
regarding the lack of phase separation near $x=0.25$, it is still
important to compare them carefully in order to assess
their descriptions of the ground states, as well as possibly
put even lower limits on the possible dopings having phase
separation.  

\begin{figure}[ht]
\epsfxsize=2.5 in\centerline{\epsffile{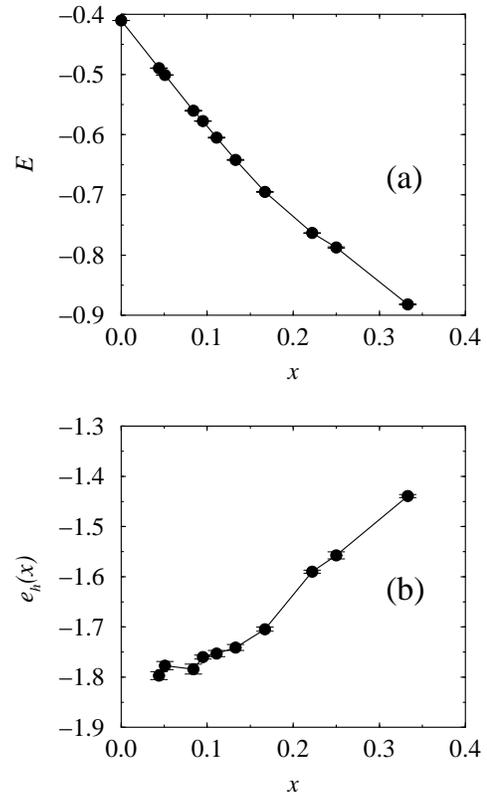}}
\caption{(a) Energy per site in $L\times6$ \tj systems, with
$J/t=0.35$ for a variety of lengths $L$ up to $L=19$.
Cylindrical boundary conditions were used, along with extrapolation
to reach zero truncation error.
(b) Energy per hole $e_h(x)$ for the same systems as in (a).
}
\end{figure}
 
In Fig. 1(a), we show DMRG results for
the energy per site, as a function of
overall doping, for $L\times6$ \tj lattices, with $J/t=0.35$
for a variety of lengths $L$ up to $L=19$. 
Note the near linearity of the data for $x \le
0.12$.  If our uncertainties were much larger, we might have
been tempted to claim phase separation based on this
data.  This near linearity of the data illustrates the numerical difficulty
of the problem---precise linearity in an exact calculation
in the thermodynamic limit is an indication of phase
separation, whereas deviations may be finite size effects,
numerical errors, or they may indicate the absence of 
phase separation. In this case, as we will discuss, the near
linearity reflects the weak repulsion of the four-hole domain walls which
form in these lattices, wrapping around the $L\times6$ cylinder.
With DMRG, we
are able to resolve the energy quite precisely and reliably
on these $L\times6$ systems.
The major constraint for DMRG is the system's width---the accuracy falls
off rapidly for wider systems. Green's function Monte Carlo can treat wider
systems, but the presence of the fermion sign problem makes the
result depend, perhaps strongly, on a trial or guiding 
wavefunction which is usually chosen to have uniform hole density. 
Thus it is essential to assess the relative importance of
finite size effects versus the approximations used to 
control the fermion sign problem.

In Fig. 1(b) we show the same data plotted as the energy per
hole $e_h(x)$ (following Ref. \cite{EKL}), relative to the undoped system:
\begin{equation}
e_h(x) = \frac{E(N_h)-E(0)}{N_h},
\end{equation}
where $N_h$ is the number of holes, and $E(N)$ is the energy of
the system with $N$ holes.
In this case, phase
separation would be seen as a minimum away from $x=0$. We see
no evidence for such a minimum, but the results are far
from conclusive for $x \le 0.1$.  At $x \sim 0.25$, however,
the results clearly indicate the absence of phase separation.
The energy per hole is about $0.25t$ higher at $x \sim 0.25$ than
at small $x$.  As we discuss below, this energy difference
is about an order of magnitude higher than typical
finite size effects at this width.
Therefore, we can quite safely conclude, in agreement with
other simulations, that there is no phase
separation near $x \sim 0.25$. 

\begin{figure}[ht]
\epsfxsize=2.5 in\centerline{\epsffile{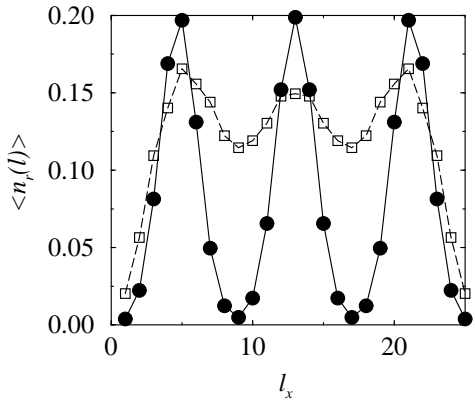}}
\caption{Density of holes per site as a function of the
$x$-coordinate $l_x$ in a $25\times6$ \tj system (filled
circles),
with 12 holes, compared with the hole density in a $25\times1$
system, with 3 holes (open squares).
}
\end{figure}

In fact, the systems shown in Fig. 1 with $x < 0.25$
are striped, with four-hole stripes wrapping around the
cylindrical systems.  In Fig. 2 we show the average hole density
per site as a function of the $x$-coordinate 
on a $25\times6$ system with 12 holes.
For comparison, so that one can judge the
effects of the open boundary conditions, we show a $25\times1$
system with 3 holes. In the case of a single chain, the
charge density oscillations decay as a power law away from the open
ends, but in the thermodynamic limit the system is uniform.
In the case of the $L\times6$ systems, however, the amplitude
of the oscillations is much larger and much more anharmonic at
this doping.
Using DMRG,
we have found evidence for
striped ground states for a wide range of dopings in the \tj
model \cite{stripe,energetics}.  Importantly, we have found that
stripes with a linear doping near 1/2 on long domain walls
are the lowest energy 
configurations at low doping.

The striped nature of the ground state tells
us why the energy shown in Fig. 1(a) is so nearly linear: 
adjacent stripes repel, but only very weakly at large distances.
The repulsion appears to be due to overlap of the hole densities
in the adjacent stripes, and falls off roughly exponentially with
the separation at short to intermediate distances.
Thus, the system at low doping becomes almost infinitely
compressible, making the energy per site as a function of
doping nearly linear, and suggesting phase separation. 

As mentioned above, our results showing the presence of 
stripes disagree with most Green's function Monte Carlo work. 
This may be because
the uniform trial wave functions used to
date in these calculations bias the calculations towards
uniform states.
Note also that all that is necessary to generate false signals
of phase separation is that one's trial wave function be 
substantially worse for low doping than for high doping.
Fortunately, it is possible to compare the various
calculations because most are variational---if 
a trial wavefunction is poor, it will produce
a higher energy result than it should. Even in cases where
the calculation is not variational, a poor calculation will
often result in an energy above the true ground state energy.
In the case of DMRG,
two results are available: a variational energy, and a more
accurate but nonvariational energy coming from extrapolating
the truncation error to zero. However, we find that the shift
in energy in DMRG coming from this extrapolation is small compared
to the differences in energies between different methods.

In Fig. 3, we compare the energies per hole from DMRG and
exact diagonalization calculations for a number of systems with $J/t=0.5$.
We see that the DMRG results agree nicely with exact
diagonalization\cite{Lanczos}, within finite size effects. Note that perhaps
the largest finite size effect in the energy per hole comes
from how the reference undoped energy $E(0)$ is defined. For the
$N=20$ and $N=26$ lattices studied with the Lanczos method,
the undoped Heisenberg system is unfrustrated, and the ground
state energy per site is lower than in an infinite system.
This results in a higher energy per hole (Lanczos-I) than if one
uses the infinite-system energy per site $-1.16944(4) J$
\cite{sandvik} as reference (Lanczos-II). This effect is
much less pronounced on the larger $N=26$ site system.
However, the
corresponding doped systems are not necessarily unfrustrated.
In particular, formation of a single stripe would be frustrated by these
boundary conditions. This would make it very difficult to draw
any conclusions about stripe stability from the Lanczos data alone.
However, the Lanczos data provide an important check on the
accuracy of the calculations on larger systems.
For the DMRG with cylindrical boundary
conditions, one cannot use the infinite-system reference energy,
so the same undoped system is used. Using the same system as
reference results in a cancelation of exchange energies associated with the
open sides, reducing the finite size effects.
Note that with cylindrical boundary
conditions, striping is not frustrated.

\begin{figure}[ht]
\epsfxsize=2.7 in\centerline{\epsffile{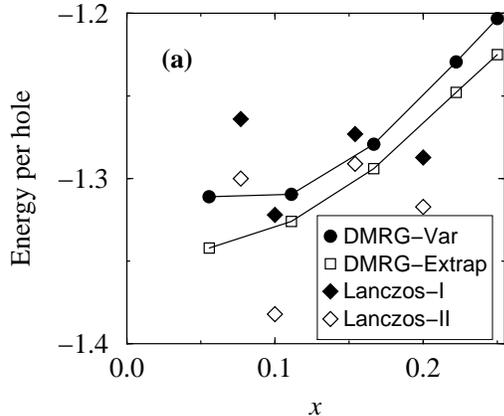}}
\caption{Energy per hole $e_h$, in units of $t$, using DMRG and
exact diagonalizations.  The points labeled DMRG-Var are from DMRG
calculations performed with cylindrical boundary conditions on 
$12\times6$ systems, and are variational. The points labeled
DMRG-Extrap are extrapolated to zero truncation error. 
Here, the energies are defined relative to the DMRG energy
of an undoped system on the same lattice. These 
calculations for undoped systems are much more accurate than for
the doped systems, and we ignore any errors in these energies in
claiming that the DMRG-Var results are variational.
Two different types of exact diagonalization results are shown,
for systems with $N=20$ and $N=26$ sites, with two or four
holes.
For the points labeled Lanczos-I, the energy is defined relative
to the exact undoped energy on the same lattice. For the
Lanczos-II points, we used as the undoped energy the energy per
site for an infinite Heisenberg lattice, multiplied by the 
number of sites.
}
\end{figure}

In Fig. 4, we compare DMRG and Green's function Monte Carlo
results.  The results of Hellberg and Manousakis 
are based on unpublished data \cite{hmdata}
which was summarized in Ref. \cite{hellberg}. The points 
shown are a representative subset of the results used in Ref. \cite{hellberg}.
A fit to all the data, showing a minimum near $x=0.14$, was
the basis for the conclusion of phase separation at $J/t=0.5$ in Ref.
\cite{hellberg}.
For the $6\times6$ system with 4 holes and the $7\times7$ system with
7 holes, the results of HM and Calandra, et.
al. \cite{sorella} are in fairly good agreement. The DMRG energy
on a $12\times6$ system is lower. We attribute this energy
difference to the energy associated with the formation of
stripes. Note that the $6\times6$ system with 4 holes would
be frustrated if the holes formed  a single stripe. To study
a similar, frustrated $6\times6$ system with DMRG, we have
applied frustrating staggered magnetic fields to the open ends
of a $6\times6$
system with cylindrical BCs. The points labeled ``AF'' have this
field, which would favor N\'eel order, but frustrate the $\pi$
phase shift of a stripe. Two field strengths were used,
$h=0.1$ (the lower energy point) and $h=0.2$. The calculation
labeled $\pi$ had a $\pi$
phase shift (with $h=0.1$) favoring a stripe in the applied staggered fields.
As reference energies, the equivalent undoped, unfrustrated 
Heisenberg system was used in all cases. Application of the
frustating fields brings the energy of the $6\times6$ system
very close to the QMC results. We interpret this to mean
that the stabilization energy of the stripe is nearly balanced
by the frustration of the boundary conditions. Thus the QMC
energy of Calandra, et.
al. \cite{sorella} on the $6\times6$ system may be quite accurate, even if it
does not  have a stripe. (We note that their results for 
the $N=26$ sites system compare well with Lanczos.)
On a $7\times6$ system with 4 holes (not shown)
Calandra, et.  al. obtain an energy per hole of about $-1.26t$,
very close to their result for the $6\times6$ system. However,
in this case, where a single stripe is not frustrated,
we believe that the energy should be about the same as in the
$12\times6$ system near this doping, namely about $-1.31t$.
In general, on $L\times6$ systems which do not
frustrate stripes, we expect the QMC results to be too high
by a stripe stabilization energy of $\sim 0.05t$ per hole 
near $x \sim 0.1$.

\begin{figure}[ht]
\epsfxsize=2.8 in\centerline{\epsffile{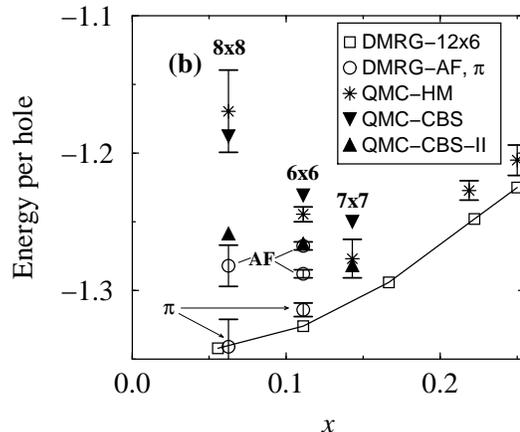}}
\caption{Energy per hole $e_h$, in units of $t$, comparing
DMRG and Green's Function Monte Carlo data.
The points labeled DMRG-12x6 are the DMRG-Extrap points from
Fig. 3.
The QMC points are from a variety of periodic lattices, some
of which are labeled. In this
case, the energies are defined relative to the identical undoped
system, as in the Lanczos-I data. However, for the $7\times7$
system, the undoped system would be frustrated, and therefore
an extrapolation using the $6\times6$ and $8\times8$ undoped systems
and the known finite size dependence on system size was used
to obtain a reference energy for this case.
The points labeled 
QMC-CBS are variational fixed node quantum Monte Carlo 
calculations provided 
to us by Calandra, Becca, and Sorella. The points labeled QMC-CBS-II
are from their stochastic reconfiguration method ($p=6$), which is
not variational.
The points labeled QMC-HM are from the calculations of Hellberg
and Manousakis, which use a released-node procedure and are
not strictly variational.
The DMRG-AF,$\pi$ calculations are described in the text.
Where not shown, error bars are smaller than the symbols.
}
\end{figure}

Measurements of the hole-hole correlation
function by Calandra, et.  al. in \cite{calandra} 
were made using the less accurate 
fixed node approximation.  Even in these
measurements, they found some signs of incommensurate
correlations indicating incipient fluctuating stripes. 

The data point for the $8\times8$ system with 4 holes of HM appears to
be anomalously high. Our results with frustrating and nonfrustrating
fields (all with $h=0.1$) on the $8\times8$ system give results very similar
to those of the $6\times6$ system, and with much lower energy
than found by HM. The stochastic reconfiguration result 
of Calandra, et. al. for the same system is also much lower.
The high energy on the $8\times8$ systems 
appears to have been important for the conclusion of phase
separation near $x=0.14$ at $J/t=0.5$ of HM. The best data of
Calandra, et. al. also show a slight minimum near $x=0.14$, but
in Ref. \cite{calandra} calculations on larger systems showed
this to be only a finite size effect.
Note that aside from the small systems studied by Lanczos,
and cases where issues of frustration arise,
typical finite size effects are rather small in the energy
per hole.  
Systems with about 50 sites were found in
\cite{calandra} to have finite size effects of about
$0.01t-0.02t$ per hole, when compared to much larger systems.
Similar finite size effects were reported in \cite{energetics}. 
These finite size effects may be important for the
determination of phase separation, but they are small enough
to allow us to compare the various methods on slightly different
lattices.
We also find that the energy
per hole is insensitive to the use of open boundary conditions
on the two short ends of our $L\times6$ systems. For example,
in comparing a $12\times6$ system and the central
$12\times6$ region of a $24\times6$ system, keeping the doping
constant, we find
a difference of less than $0.01t$ in the energy per hole. 

A recent analysis \cite{casimir} of Casimir forces involving
spin-wave modes has found that in the limit of low doping, in
the absence of Coulomb interactions, static stripes attract with
an interaction decaying as $r^{-2}$. One can also estimate the
coefficients in front of the leading decay terms; for stripes,
the behavior is roughly $10^{-2}J r^{-2}$ per unit length, with
$r$ in lattice spacings, for large $r$. 
At all length scales this force is a small 
correction to Coulomb interactions,
which decay as $r^{-1}$ with a larger coefficient, assuming
dielectric screening. This means that the Casimir effect cannot
have a role in frustrated phase separation.
However, the Casimir force is potentially relevant to
the issue of phase separation in the pure \tj model, since
it would induce an attraction between stripes, causing
an unusual form of true macroscopic phase separation into
regions having widely spaced stripes, and regions without
stripes \cite{kivelson}.  It is important to estimate at what dopings this 
force can come into play.  

Our simulations automatically include the Casimir
effects as well as other short range effects.
At distances between stripes in our $L\times6$ systems of 
up to about 10-12 lattice spacings, we have found only pure repulsion.
We believe this is because the wavefunctions
of the holes in the stripes extend beyond the stripe in 
exponentially decaying tails, and the overlap of these
tails apparently causes higher hole kinetic energy.
At larger separations the energies are too small to resolve.
This distance puts a limit on the maximum doping, for the Casimir
effect to be important, of about 0.06.
We have also fit the short range repulsion in $L\times6$ 
striped systems to an exponential form \cite{energetics}; 
we find the potential is roughly $0.6t \exp(-r/1.8)$ per
unit length. If this is repulsion is assumed, the Casimir
effect becomes dominant at distances of about $r=20$ between
stripes, corresponding to dopings of less than 0.025.
The temperature at which such small energies could be relevant
would be less than 1K, assuming $J \approx 1500K$.
Despite its limited applicable doping range, 
the Casimir effect illustrates the extreme
difficulty of resolving the issue of phase separation
in the \tj model in the limit of small doping, using
only numerical simulations. However, this question is
of very limited relevance physically.

These comparisons indicate that the DMRG calculations are
quite reliable, at least in terms of the ground state energies.
Based on this, we conclude that the short distance
behavior and correlations of these systems, which affect the
energy most strongly, are reliably determined by DMRG.
However, we would like to address the question of the boundary
conditions used by DMRG in somewhat more detail. It has
been suggested that the stripes we see with DMRG may be due
to the use of cylindrical boundary conditions, that they are
artifacts which would not appear in ``more realistic''
periodic boundary conditions. We disagree with this position.
In using finite size clusters
to study models which may have broken-symmetry ground states,
one often introduces a symmetry-breaking field and
then studies the limiting behavior by first letting the size of
the system go to infinity and then letting the strength of the
perturbation go to zero. We view the open end boundary
conditions in the DMRG calculations in this way. Far from
being artificial, they are important for understanding the
physics. Of course, at present we are unable to carry out a
proper finite-size scaling analysis to obtain the infinite size
limit. Such a study would require very large lattices, since
the domain wall spacing rather than the lattice spacing sets
the lattice sizes required. Nevertheless, we have compared on
numerous occasions systems of different lengths, and not seen
significant reduction in the stripe amplitudes. Furthermore,
while we have seen various arrangements of stripes depending
on boundary conditions, we have been {\it unable} to stabilize
any uniform states. In contrast, we {\it are}  able to observe
an essentially uniform ground
state even with open boundary conditions; they occur when
a next-nearest neighbor hopping $t'$ is made large
enough ($t' \sim 0.3 t$) \cite{tprime}. The effect of this term
is to destabilize the domain walls \cite{tohyama} and favor a gas of pairs.

\begin{figure}[ht]
\epsfxsize=3.0 in\centerline{\epsffile{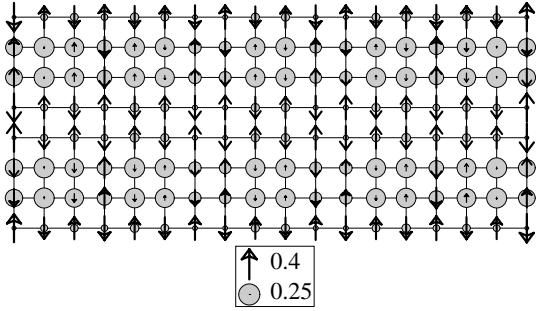}}
\caption{Hole density and spin moments showing longitudinal
stripes on a 
$18\times 8$  \tj lattice with cylindrical boundary conditions,
$J/t=0.35$, and 20 holes.
The diameter of the circles
is proportional to the hole density $1-\langle n_i\rangle$ on
the $i^{\rm th}$ site and the length of the arrows is
proportional to $\langle S^z_i\rangle$, according to the scales
shown. Differently styled arrows are used to show the two different
antiferromagetic domains. This structure
depends on the boundary conditions as discussed in the text.
}
\end{figure}

As an illustration of the robustness of the striped state,
we have made an effort to stabilize {\it longitudinal}  half-filled
stripes in
$L\times8$ systems with cylindrical boundary conditions. 
This is somewhat difficult, because the transverse stripes
appear to have slightly lower energy. Furthermore, we have found that
domain walls do not like to end on open boundaries,
which seem to repel wall ends. However, we have found that we can
stabilize the ends nicely by increasing the hopping slightly
on a single edge link at which we wish the domain wall to end. In Fig. 5
we show the hole and spin densities in an $18\times8$ system
with two longitudinal stripes. We used a hopping of $1.2t$ on
the second and sixth vertical edge links on both the left and
right edges, and we also applied staggered fields with a $\pi$
phase shift built in on sites (1,1), (1,4), (1,5), and (1,8),
and the equivalent sites on the right edge.
To stabilize the stripe configuration it was also necessary to
apply pinning fields throughout the system during the warmup sweep
and first several finite system sweeps. Because of the mapping
of the sites in the 2D system onto an effective 1D chain in
DMRG, during these first
sweeps the system is much better equilibrated in the $y$ direction than
the $x$-direction, and the system is unstable to the formation of transverse
stripes.  After these sweeps, all
of the interior fields were turned off and about a dozen more
sweeps were performed, with the final number of states kept
per block equal to 1600. This calculation shows that with
pinning terms applied only at the edges, a rather long cylinder
supports longitudinal stripes. These stripes cannot be regarded
as simple charge density oscillations induced by boundaries, as
one could argue one has in the single chain system shown in Fig. 2. 
Of course, on a long
system the state with longitudinal stripes might have higher
energy than a state with the bulk having transverse stripes.
DMRG is unable to tunnel between two states which differ so
much over large length scales. However, we believe that DMRG
would have no trouble making the system shown in Fig. 5 uniform
if the correct ground state was uniform, simply by smearing out
the stripes in the central region.

Thus, we believe our results 
imply that the pure 2D \tj model, in the small-$J/t$ regime
most relevant to the cuprates, and with dopings near $x \sim
0.1$, has a ground state which is striped \cite{ref}.
By this, we mean that dynamical spin and charge susceptibility 
measurements will show either sharp peaks or divergences
characteristic of dynamic or static stripes, respectively. 
Furthermore, we believe that there are no low-lying states which
do not have some signs of static or dynamic stripes.  
Specifically, for $J/t=0.5$ we estimate the lowest energy
stripeless states are about $0.05t$ per hole higher in energy
than the ground state.
We believe that if one tries to write down 
variational wavefunctions for
the ground state, and omits the striping behavior, one will
not achieve a low energy state, even in cases where
the stripes are purely dynamic. 
Finally, the energy scales involved
suggest that a proper description of superconductivity
requires taking these stripes into account.

We thank   T.~Tohyama, M.~Calandra, F.~Becca, D.~Hone, C.~S.~Hellberg,
D. Poilblanc,
and especially E.~Dagotto, E. Manousakis, and S.~Sorella for
stimulating conversations and for providing us with results
from their calculations.
S.R.~White acknowledges support from the NSF under
grant \#DMR98-70930 and D.J.~Scalapino acknowledges support
from the NSF
under grant \#DMR95-27304.

%



%
\newpage


\begin{references}
\bibitem{ref} We consider it to be a separate issue whether the
stripes are
static or dynamic.  In our calculations, the stripes are static.
In particular, at low doping, the half-filled stripes show
no signs of significant fluctuations. On the other hand, at
higher dopings, we find some signs that filled stripes may
fluctuate \cite{energetics}.
However, even at low doping, it would require a very careful finite
size scaling analysis on very large systems to rule out low energy
dynamic quantum fluctuations.
\end{references}
\end{document}